\documentclass[10pt, oneside, reqno]{amsart}
\input ./math_headers_amsart.sty

\usepackage[backend=biber,
            isbn=false,
            doi=false,
            maxbibnames=5,
            style=alphabetic,
            citestyle=alphabetic]{biblatex}

\usepackage{scalerel}

\newcommand\reallywidehat[1]{\arraycolsep=0pt\relax%
\begin{array}{c}
\stretchto{
  \scaleto{
    \scalerel*[\widthof{\ensuremath{#1}}]{\kern-.5pt\bigwedge\kern-.5pt}
    {\rule[-\textheight/2]{1ex}{\textheight}} 
  }{\textheight} %
}{0.5ex}\\           
#1\\                 
\rule{-1ex}{0ex}
\end{array}
}
\bibliography{defect}

\title{Defects via factorization algebras}
\author{Ivan Contreras \and Chris Elliott \and Owen Gwilliam}
\date{\today}

\begin{document}

\begin{abstract}
We provide a mathematical formulation of the idea of a defect for a field theory, in terms of the factorization algebra of observables and using the BV formalism.  
Our approach follows a well-known ansatz identifying a defect as a boundary condition along the boundary of a blow-up, 
but it uses recent work of Butson--Yoo and Rabinovich on boundary conditions and their associated factorization algebras to implement the ansatz.  
We describe how a range of natural examples of defects fits into our framework.
\end{abstract}

\maketitle

\section{Introduction}

The notion of a defect in a field theory has played an increasingly important role in physics and, perhaps surprisingly, in mathematics.
Our goal in this short paper is to offer a mathematical formulation of defects in field theory
that builds upon recent progress with the Batalin-Vilkovisky (BV) formalism and higher algebra.
Let us start by outlining one interpretation of the idea of a defect.

Suppose we are studying a field theory $\mathcal{T}$ on a manifold $M$. 
Loosely speaking, a {\em defect} is a modification of the theory along a submanifold $D \subset M$ in a way that produces a new field theory.
We call $D$ the {\em support} of the defect,
and the dimension of the defect is the dimension of $D$.
Well-known examples include the Wilson and `t Hooft line defects in Yang--Mills theory:
these are supported along one-dimensional submanifolds of a four-dimensional manifold.
(In physics these defects provide a means to characterize the phase of a gauge theory, such as confining or Coulombic.)

We thus view a defect as consisting of a pair of data: the submanifold on which its supported and how it affects (or couples) to the ambient theory.

\begin{remark}
Note that this approach diverges from the terminology in some communities,
which might prefer to view a defect (or extended operator) as defined for some large class of submanifolds.
For instance, people often refer to a representation $V$ of a group $G$ as giving a Wilson line operator, since one can produce a line defect from $V$ on a large class of embedded 1-manifolds.
\end{remark}

There is an ansatz for how to produce defects that will guide our approach in this paper;
it works by reducing the problem to studying boundary conditions.

\begin{ansatz}
Let $\mathcal{T}$ be a classical field theory on a manifold $M$, 
which determines a system of partial differential equations. 
There are two steps to making a defect along a submanifold $D \subset M$:
\begin{enumerate}
    \item Take a small tubular neighborhood $\widetilde{D} \supset D$ whose complement is a manifold with boundary $M \setminus \widetilde{D}$.
    \item Specify a boundary condition for the theory $\mathcal{T}$ on this manifold with boundary.
\end{enumerate}
If one takes a small enough neighborhood (or works in some limit as the neighborhood shrinks), 
then this new theory (with the imposed boundary condition) looks like the original theory but modified along~$D$, as the boundary condition affects the solutions to the equations of motion.
We call this new theory the {\em theory $\mathcal{T}$ with a defect along~$D$.}
\end{ansatz}

If we specify a quantization of $\mathcal{T}$ on $M \setminus \widetilde{D}$ with this boundary condition,
then we produce a defect for a quantization of~$\cT$.
See Remark~\ref{rmk on history} for some background about this ansatz.

\begin{remark}
Note that after removing $D$ from~$M$, there may be solutions to the equations of motion that do not extend across~$D$. 
More generally, one may even consider theories $\cT$ defined only on $M - D$.  
We will continue to use the term ``defect'' for such situations, where the boundary condition for the theory $\cT$ on $M \setminus \widetilde{D}$ prescribes the allowed limiting behavior of $\cT$ as one approaches~$D$.  
This situation is also commonly referred to by the term \emph{disorder operator}.
(For a discussion of {\em order} operators, see Remark~\ref{rmk: order ops} below.)
\end{remark}

An appealing aspect of this ansatz is that it is constructive, 
at least once one spells out how to produce boundary conditions for quantum theories.
A drawback is that it is not a definition: certainly given any defect along $D$ and any tubular neighborhood $\wt D$ we should be able to obtain a boundary condition as described in the ansatz, but we will see that this assignment generally does not need to be an equivalence, because of the reliance within the ansatz of a specific choice of neighborhood $\wt D$.  

A slight shift of perspective, however, offers a useful setting to pose an actual definition.
We shift our focus to the algebra of observables of a theory with defect, rather than focussing directly on the theory itself, 
much as it is often fruitful to focus on the algebra of functions on a space rather than directly on the space.
In the setting of QFT (at least in the perturbative regime), the observables form a {\em factorization} algebra on the spacetime manifold \cite{CG2}.
Recall that a factorization algebra $\cA$ on a manifold $M$ is a local-to-global object,
akin to a sheaf. (For a systematic treatment, see \cite{CG1, AF} and references therein.)
In particular, it determines a functor $\cA \colon \mr{Open}(M) \to \mr{Ch}$ assigning a cochain complex $\cA(U)$ to each open set $U \subset M$.
For a QFT, $\cA(U)$ consists of the observables with support in $U$, i.e., that depend on the behavior of the fields only in the region $U$.
The key result of \cite{CG2} is that if one constructs the QFT in the BV formalism, 
the complex of observables forms a factorization algebra.

In the language of factorization algebras, we proffer the following view on defects.

\begin{definition}
\label{definition: fact def}
Let $M$ be a manifold and $D \subset M$ a submanifold.
Let $\cA$ be a factorization algebra on a manifold $M-D$.
A {\em defect along $D$ for $\cA$} is a factorization algebra $\cB$ on $M$ with an isomorphism
\begin{equation}
\label{eq: fact def}
\phi \colon \cB|_{M - D} \xto{\cong} \cA
\end{equation}
on the complement of $D$.
\end{definition}

We view $\cB$ as an extension of $\cA$ along $D$, with specifically prescribed modified behavior.
In other words, we view $M$ as stratified $D \subset M$ and we ask for factorization algebras that agree with $\cA$ on the big stratum~$M - D$. 
When $\cA$ is the algebra of observables for a field theory,
this definition clearly matches the behavior of the observables of a defect in its heuristic form.

Our paper explains how one can implement the ansatz carefully to produce such factorization algebras modelling a defect.
More precisely, given a classical BV field theory $\cT$ on a manifold $M$, we will produce defects for $\obscl_\cT$ in the sense of Definition~\ref{definition: fact def}.
The construction is outlined in Section~\ref{sec: con}.
The factorization algebra modelling the defect depends explicitly on the choice of tubular neighborhood of~$D$.

When, however, the theory satisfies a useful property near the defect --- it is {\em topological normal to $D$} --- we can then use recent results of Rabinovich \cite{Rab20,RabinovichThesis} to construct defects that do not depend on the size of the neighborhood.
Moreover, Rabinovich's results allow one to construct the quantum observables for the defects, when a BV quantization exists.
We discuss these results in Section~\ref{sec: theorem}.

Finally, we describe a number of examples, organized by codimension of the defect,
such as magnetic monopoles and Wilson lines, among others.

\subsection{Some history, context, and future directions}
\label{rmk on history}

The ansatz we have discussed is well-known, at least among quantum field theorists, and has probably been known for several decades.  
We are not knowledgeable, however, of how this idea appeared and evolved within the literature. 
We found an explicit and useful articulation in~\cite{Kapustin} but it begins by acknowledging the idea is well-known.
(Insights and suggestions about the history and literature are welcome.)

The essential idea about how to capture defects with factorization algebras has also  been floating around the community of factorization algebraists for at least a decade,
and hence some version was known by many people.
What prompted us to document and extend these ideas is the powerful work of Rabinovich,
which makes it possible to implement the ansatz precisely and rigorously using factorization algebras in a broad class of examples. 

A careful treatment of ``defects for a factorization algebra'' is already available in the topological setting, thanks to the pioneering work of Ayala--Francis--Tanaka.
The reader is encouraged to explore Section 4.3 of~\cite{AyalaFrancisTanaka} for the complete story, 
but here we will gloss the key results and explain how they fit into the story of this paper.
(We will suppress all subtleties about framings.
Moreover, that paper works in the setting of $\infty$-categories; below we will simply use the term ``category.'' 
The interested reader can find details in~\cite{AyalaFrancisTanaka}.)

Recall that a factorization algebra $\cF$ on an $n$-dimensional manifold is {\em locally constant} if the structure map $\cF(U) \to \cF(U')$ is an equivalence whenever $U \subset U'$ are each open subsets diffeomorphic to $\RR^n$. 
In any open ball in the manifold, such a factorization algebra is described by an $E_n$-algebra.
Ayala--Francis--Tanaka offer a characterization of all the ways to extend a locally constant factorization algebra $\cA$ on $M - D$ to a factorization algebra $\cB$ on $M$ that is locally constant when restricted along $D$.
That means that for any inclusion of disks $U \subset U'$ where $U \cap D$ and $U' \cap D$ are nonempty,
the map $\cB(U) \to \cB(U')$ is an equivalence;
hence $\cB$ along $D$ is locally determined by an $E_k$-algebra if $k$ is the dimension of~$D$.
Hence, the local situation is when $M = \RR^n$ and $D = \RR^k$ is a vector subspace.
A key result (Proposition 4.8) of \cite{AyalaFrancisTanaka} is that given the ``bulk'' $E_n$-algebra $A$, which determines $\cA$ on $\RR^n -\RR^k$, 
we need to pick an $E_k$-algebra $B$ with values in the category of left modules for $\int_{S^{n-k-1}} A$,
where we view this $n-k-1$-sphere as linking the defect.
This factorization homology $\int_{S^{n-k-1}} A$ encodes how $A$ has to act on $B$.
(Alternatively, we pick a Swiss cheese algebra of dimensions $(k+1,k)$ of the form $(\int_{S^{n-k-1}} A,B)$.)
Ayala--Francis--Tanaka explain how to extend factorization homology for such pairs $(A,B)$ to obtain all factorization algebras that are stratified as $D^k \subset M^n$.
In other words, they reduce the classification of such defective factorization algebras to classifying Swiss cheese algebras.

This result is powerful and satisfying, and it tells one how to use computational methods and results from topology and algebra to classify defects (for factorization algebras).
It is wholly complementary to the physical point of view.
On the other hand, it is only applicable to topological field theories and topological defects therein;
it would be nontrivial to formulate a version that works in more geometric settings.
(It does proffer guidance, however, and it matches the behavior of, say, vertex modules corresponding to modules over a kind of ``algebra of modes'' for a vertex algebra.)
Their classification does not capture all physical expectations and requirements, however, as it merely asks for ways to extend a factorization algebra.
Physics often suggests some additional properties or features of the extension.
Compare, for instance, with the problem of identifying which modules over a deformation quantization of a Poisson algebra should be seen as physically relevant: 
not all $D$-modules seem to have a place in quantum mechanics (or at least an obvious place). 

Looking to the future, we expect the approach to defects that we describe here to generalize to several more sophisticated contexts.  
Here are two natural extensions.
\begin{enumerate}
 \item The definitions we introduce here make sense even when $D \sub M$ is not a submanifold.  All we will need is that $D$ admits an open neighborhood $U$ in $M$ with a continuous contracting map $U \to D$.  For example, one could consider situations where the defect space $D$ is singular, like a nice immersion.
 \item Suppose $D$ is a stratified manifold (possibly singular, as in the example above).  One could form an extension of the approach explained here to model theories admitting defects within defects.  We hope that a careful analysis of this idea would lead to an instantiation of higher categorical structures associated to defects, as explained in Kapustin's ICM address~\cite{KapustinICM}.
\end{enumerate}
Pursuing these directions would allow for many more physical ideas to be translated into the language of factorization algebras.

\begin{remark}
The approach to defects that we take here is related to Costello and Li's conjectures \cite{CostelloLi} about holography for twisted supersymmetric theories,
which uses Koszul duality to understand and characterize defects. 
In their work on this approach, Paquette and Williams \cite{PaqWil} develop language for the consideration of line defects, and in particular of a \emph{universal} defect.  
Their definition provides examples of line defects in the sense of this paper.  
More specifically, their line defects can be thought of as ``order'' type, and they work under the assumption that the theory is topological in the direction spanned by the line.  
We will describe some examples of this type in Section~\ref{sec: ex}.
\end{remark}

\subsection*{Acknowledgments}

We have benefited from discussions with many people about defects, factorization algebras, and the BV formalism.
We would like to thank Iv\'an Burbano, Dylan Butson, Alberto Cattaneo, Kevin Costello, John Francis, Ben Heidenreich,  Rune Haugseng, John Huerta, Theo Johnson-Freyd, Pavel Mnev, Eugene Rabinovich, Ingmar Saberi, Pavel Safronov, Claudia Scheimbauer, Michele Schiavina, Christoph Schweigert, Stephan Stolz, Matt Szczesny, Peter Teichner, Alessandro Valentino, Konstantin Wernli, Brian Williams, and Philsang Yoo;
undoubtedly more should be listed, as this is a frequent topic of conversation.
The National Science Foundation supported O.G. through DMS Grants No. 1812049 and 2042052. I.C. thanks the Amherst College Provost and Dean of the Faculty's Research Fellowship (2021-2022).

\section{The construction}
\label{sec: con}

Let $M$ be a smooth manifold and let $\cT$ denote a classical BV theory on~$M$, 
following Costello's definition in Chapter~5 of~\cite{CosBook}.
The associated factorization algebra $\obscl_\cT$ is described in Chapter~5 of~\cite{CG2}.
We will now explain how to produce something analogous for theories with defects.

Let $D \subset M$ be a submanifold.
Equip the restriction of the tangent bundle $TM|_D$ with a fiberwise metric,
and let $n \colon N \to D$ denote the normal bundle to $D$, 
which inherits a fiberwise metric.
Let $B_r(D) \subset N$ denote the disk bundle of radius less than~$r$,
and suppose that the exponential map provides a diffeomorphism
\[
\rho \colon B_3(D) \xto{\cong} U
\]
with a tubular neighborhood $U$ of $D$.
(We can always adjust the metric on $TM|_D$ to accomplish this.)
We then have a family of tubular neighborhoods
\[
U_t = \rho(B_t(D)) 
\]
around $D$ parametrized by $t \in (0,3)$.
Note that we will only work with values of $t$ up to~$1$.

Let $M_t = M - U_t$ denote the complement of the tubular neighborhood $U_t$.
It is a manifold with boundary,
and let 
\[
\partial_D M_t = \partial U_t \subset M_t
\]
denote its {\em boundary along $D$}.
(If $M$ has boundary itself, we will ignore that region and focus only on the boundary introduced by excising a neighborhood of $D$.)
Suppose we have a theory $\cT$ defined on $M - D$, 
so that the theory $\cT$ restricts to a theory on $M_t$,
and, following the ansatz, we wish to impose a boundary condition $\cL_t$ along~$\partial_D M_t$
and construct the factorization algebra of classical observables given that boundary condition. 
We will use $\obscl_{\cL_t}$ to denote this factorization algebra on~$M_t$.  (In a moment we will describe what we mean by a boundary condition and how it determines a factorization algebra.)

Suppose we have produced this factorization algebra~$\obscl_{\cL_t}$ and want to produce the associated factorization algebra modelling the observables in the theory on $M$ with defect. 
The essential idea is simple: provide a map $\pi_t \colon M_t \to M$ collapsing $\partial_D M_t$ to $D$ and push $\obscl_\cL$ forward along this map.
This pushforward $(\pi_t)_* \obscl_{\cL_t}$ ought to model a defect for~$\obscl_\cT$ on~$M$.
We now make more precise what we want to do.

For a specified value of $t$ in $(0,1)$, let $f_t \colon [0,3] \to [0,3]$ be a smooth non-decreasing function such that 
\begin{align}
\label{f specs}
\begin{split}
 &f_t \text{ is a diffeomorphism preserving the boundary},\\
 &f_t(s) = 0 \text{ for } s \in [0,t] \text{ and,}\\
 &f_t(s) = s \text{ for } s \in [2t,3].
\end{split}
\end{align}
Such a function exists, and the space of these functions is contractible.
We now use it to define a map $F_t: B_3(D) \to B_3(D)$ as follows.
For each point $x \in D$, let $B_3(D)_x$ denote the fiber over $x$ in the disk bundle, 
and let $|v|$ denote the length of a point $v \in B_3(D)_x$.
Then set $F_t(x,v) = (x, f_t(|v|)v/|v|)$ for any point $(x,v) \in B_3(D)$.
This map rescales each fiber so that vectors within distance $t$ of $x$ get collapsed to $x$,
those with distance at least $2t$ are left unchanged, 
and those in between get stretched.

Now define
\[
\pi_t(p) = 
\begin{cases} 
F_t \circ \rho^{-1}(x), & p \in U \\ 
p, & p \in M - U_{2t} = M_{2t}
\end{cases}.
\]
Note that outside $U_{2t}$, the map $\pi_t$ leaves $M$ unchanged,
while inside it collapses the boundary $\partial_D M_t$ onto $D$ and ``stretches'' the region between $\partial_D M_t$ and $\partial_D M_{2t}$ to fill the neighborhood of~$D$.

Hence, given any factorization algebra $\cA$ on $M$, we find that
\[
\cA|_{M_{2t}} \cong \left( (\pi_t)_* (\cA|_{M_t}) \right)|_{M_{2t}},
\]
i.e., this construction leaves a factorization algebra unchanged outside~$U_{2t}$.  Pursuing this idea, we can define a factorization algebra that models observables on a theory with defect well outside the neighborhood $U_{2t}$, by the following procedure.

\begin{definition}
\label{definition: constr}
We define a factorization algebra $\obscl_{t}$ --- the \emph{observables with an effective defect} --- on a manifold $M$ with submanifold $D \subset M$ using the following input data:
\begin{enumerate}
\item a classical BV field theory $\cT$ on the manifold~$M-D$, 
\item a tubular neighborhood $U$ of $M-D$ with collar coordinate $\rho \colon B_3(D) \cong U$,
\item a choice of real number $t$ in $(0,1)$,
\item a function $f_t$ satisfying conditions~\eqref{f specs},
\item a local boundary condition $\cL_t$ along~$\partial_D M_t$. 
\end{enumerate}
This data determines a factorization algebra~$\obscl_{\cL_t}$ on~$M_{t}$.
Let $\obscl_t$ denote the factorization algebra 
\[
(\pi_t)_* \obscl_{\cL_t}
\]
on~$M$.
\end{definition}

This factorization algebra $\obscl_t$ is nearly, but not exactly, a defect of $\obscl_\cT$ in the sense of Definition~\ref{definition: fact def}.
It satisfies the weaker condition that
\begin{equation}
\label{eq: eff def}
\obscl_t|_{M_{2t}} = \obscl_\cT|_{M_{2t}}.
\end{equation}
This behavior is, however, a good match with physical intuition:
it says that near $D$, the physical defect does change the behavior of the observables, 
but at some distance away from $D$  (i.e., on $M_{2t}$), the local behavior of observables are unchanged.
This physical defect changes, of course, the observables on any open set containing~$D$.
In particular, the global observables with defect $\obscl_t(M)$ are (typically) sensitive to the physical defect and do not agree with the global observables $\obscl_\cT(M)$ of the defect-free theory.

\begin{takeaway}
This construction realizes the ansatz at the level of factorization algebras,
provided one can construct the factorization algebra of observables for a classical BV theory with a local boundary condition.
\end{takeaway}

In particular, this construction produces a factorization algebra that is ``effectively'' a defect for $\obscl_\cT$ in the sense that it satisfies the weaker condition~\eqref{eq: eff def} rather than the stronger condition~\eqref{eq: fact def}.

\begin{remark} \label{limit_remark}
We could ask for slightly more data.  Instead of {\em fixing} the value of the radius $t \in (0,1)$ and then choosing $f_t, \cL_t$ for this fixed radius, 
we could choose a compatible family of such data for all values of~$t$.  
First, choose a smooth function $f(t,s) \colon (0,1] \times [0,3] \to [0,3]$ so that $f_{t_0}(s) = f(t_0,s)$ satisfies conditions~\eqref{f specs} for each choice of $t = t_0$, and $f(t, s_0)$ is monotonic for each fixed $s_0$.  
Next, define a smoothly varying family $\cL_t$ of boundary conditions for each $t \in (0,1]$.  
In this way, we produce a \emph{family} of factorization algebras $\obscl_t$ parameterized by the interval $(0,1)_t$.  One can then investigate the limiting prefactorization algebra as $t \to 0$.
\end{remark}

\subsection{On boundary conditions} \label{BC_section}

We now return to discussing what boundary conditions mean for a classical BV field theory.
A complete treatment would be lengthy and technical, so we sketch the key ideas here and point the interested reader to \cite{Rab20, RabinovichThesis} for a special class of theories of high relevance to this paper.  In addition, we will describe a number of examples in Section~\ref{sec: ex} below that we hope will give the reader a good sense of what we mean.  

\begin{remark}
We would like to highlight another powerful approach to boundary problems involving the BV formalism: the BV--BFV formalism developed by Cattaneo, Mnev, Reshetikhin, and others \cite{CMR1, CMR2}. This approach has proven to be successful in modifying BV field theory on manifolds with boundary and corners. In particular, there has been work on the BV description of AKSZ observables \cite{Mnev}, and more recently \cite{MnevSchiavinaWernli}, an interpretation of the WZW action functional in arbitrary codimension, via Witten's descent. 
\end{remark}

For a theory $\cT$ on a manifold $N$ with boundary, let $\Sol_\cT$ denote the sheaf of solutions to the equations of motion for the theory.
In the BV formalism for perturbative theories, this is a sheaf of formal derived spaces with a local $-1$-symplectic pairing, in the following sense.

\begin{definition}
Let $L$ denote a local $L_\infty$-algebra on $N$, as defined in \cite[\S 3.1.3]{CG2}, and let $BL$ denote the associated sheaf of formal derived spaces.  A \emph{local $k$-shifted symplectic structure} on $BL$ is a fiberwise nondegenerate density-valued graded skew-symmetric pairing
\[\omega \colon L \otimes L \to \mr{Dens}_N[k],\]
inducing an invariant pairing on the $L_\infty$ algebra of compactly supported sections of $L$ under integration.
\end{definition}

Near the boundary $\partial N$ (or along any hypersurface, really), we anticipate the following structure to hold in good cases.

\begin{hypothesis} \label{symplectic_hypothesis}
The formal derived space $\Sol_{\cT}^{\wedge}(\partial N)$ of jets of solutions near the boundary $\partial N$ is represented by a local $L_\infty$-algebra $L_\partial$ on $\partial N$ carrying a natural $0$-shifted local symplectic structure
\[\omega_\partial \colon L_\partial \otimes L_\partial \to \mr{Dens}_{\partial N}.\]
\end{hypothesis}

As we will see in Section~\ref{sec: ex}, this hypothesis is automatically satisfied in many natural examples, including topological theories such as BF and Chern--Simons theory, as well as non-topological examples such as Yang--Mills theory.  It would be interesting to establish the technical assumptions on the theory $\cT$ to guarantee the validity of this hypothesis, but we will not pursue this in the present paper. (In general some version of presymplectic reduction may be necessary.)

\begin{example}
We mention two quick examples to orient the reader. 
\begin{enumerate}
\item First, consider a one-dimensional field theory of maps from a line into a target Riemannian manifold $X$.
If one uses the standard action functional, then for a hypersurface (i.e., point) $t$ in the line, the space of jets of solutions is equivalent to~$T^* X$.

\item Second, consider abelian Chern-Simons theory on an oriented 3-manifold $N$,
so the sheaf of solutions is modeled by the shifted de Rham complex $\Omega^\bullet_N[1]$,
where the shift appears so that 1-forms (i.e., deformations of the flat connection $\d_{\mr{dR}}$) are in degree 0.
Pick a closed, oriented 2-dimensional hypersurface $S \subset N$.
Then jets of solutions near $S$ is equivalent to $\Omega^\bullet(S)[1]$,
as the Poincar\'e lemma tells us that the de Rham complex is insensitive (at the level of cohomology) to the normal direction.
Poincar\'e duality tells us that $\Omega^\bullet(S)[1]$ has a natural symplectic structure by the wedge-and-integrate pairing.
As a sheaf on $S$, this de Rham complex has {\em local} symplectic pairing given by the wedge product, so the local symplectic structure is well-defined even when $S$ is noncompact.
\end{enumerate}
\end{example}

We can now offer a structural formulation of a local boundary condition.

\begin{definition}
A {\em local boundary condition} for $\cT$ on $\partial N$ is a sheaf $\cL$ on $\partial N$ of formal derived spaces equipped with the structure of a Lagrangian for the sheaf $\Sol_\cT^{\wedge}$ of local $0$-shifted symplectic formal derived spaces.
\end{definition} 


Let us briefly unpack exactly what sort of data we are specifying. 
A {\em Lagrangian structure} is a map of sheaves of formal derived spaces
\[
f \colon \cL \to \Sol_\cT^{\wedge}
\]
together with a trivialization of $f^*\omega$, the pullback of the symplectic pairing, that is non-degenerate.  Non-degeneracy here means that the induced map 
\[T_{\cL} \to \mr{Fib}(T^*_{\cL} \overset {f^*\omega} \to f^*T_{\Sol_\cT^{\wedge}})\]
is an equivalence.  
The original definition in the derived setting is due to Pantev, To\"en, Vaqui\'e and Vezzosi \cite{PTVV}, and a more expository account is provided the lectures \cite{CalaqueLectures} of Calaque.
(In the setting of field theory, there are often serious functional analysis issues that appear.
See \cite{RabinovichThesis} for a careful treatment in the most important situations for this paper.)

We call this data a boundary condition because we will use it to specify which solutions we care about: we want solutions on $N$ that live in $\cL$ near the boundary.
We call this data a {\em local} boundary condition because it is a sheaf on $\partial N$,
and hence is local-to-global.

\begin{example}
We return to our examples above.
\begin{enumerate}
\item For a particle traveling through $X$, a boundary condition is precisely a Lagrangian submanifold of $T^*X$.

\item For abelian Chern-Simons theory on an oriented 3-manifold $N$ with boundary, one can specify a boundary condition on $S = \partial N$ by fixing a complex structure.
The local boundary condition is then the map of sheaves
\[\Omega^{1,\bullet}_S \hookrightarrow \Omega^\bullet_S,\]
which is manifestly isotropic because any two $(1,*)$-forms have trivial wedge product.
\end{enumerate}
\end{example}

The data of a local boundary condition determines a new sheaf on $N$ of formal derived spaces, as follows.
Let $i \colon \partial N \hookrightarrow N$ be the inclusion of the boundary.
By definition, we have a diagram of sheaves
\[
\begin{tikzcd}
  & \Sol_\cT \arrow{d}{r} \\
i_*\cL \arrow{r}{i_*f} & i_* \Sol_\cT^{\wedge}
\end{tikzcd}
\]
where $r$ is the map that restricts a solution to its jet of a solution near the boundary.
The homotopy fiber product $\cF$ of this diagram is the sheaf encoding solutions to the equations of motion that satisfy the boundary condition.
As $\cF$ is a sheaf of formal derived spaces, its algebra $\cO(\cF)$ of functions determines a cosheaf of dg commutative algebras (and hence a commutative factorization algebra) on~$N$.

\section{A theorem}
\label{sec: theorem}

We now describe a condition under which we obtain a defect in the sense of Definition~\ref{definition: fact def} because the theory has nice behavior near~$D$.
We begin by reviewing some relevant recent results of Rabinovich,
starting with classical theories before discussing quantum theories.
Boundary conditions in the classical BV formalism were previously discussed in a paper of Butson and Yoo \cite{ButsonYoo}, which also offers a bounty of examples.

For a manifold $N$ with boundary, Rabinovich \cite{Rab20} defines a classical BV theory that is {\em topological normal to its boundary} in Definition~2.4 of {\it loc. cit.} (after Butson--Yoo \cite[Definition 3.8]{ButsonYoo}).
Definition~2.21 then offers a clean characterization of a {\em local boundary condition} for such a theory,
and in Definition~2.25 he characterizes the sheaf of solutions satisfying the boundary condition.
The totality of such data he dubs a {\em classical bulk-boundary system}. 
A key result of Rabinovich's paper, explained and proved in Section~4 of {\it loc. cit.}, is that the observables of this bulk-boundary system form a factorization algebra with a defect along the boundary (in the sense of our Definition~\ref{definition: fact def}).

Let us briefly sketch Rabinovich's definition of a theory that is topological normal to its boundary.  We refer to the original paper for a full account.

\begin{definition}[See {\cite[Definition 2.4]{Rab20}}]
A classical BV theory $\cT$ on a manifold $M$ with boundary $\partial M$ is \emph{topological normal to its boundary} if there exists a tubular neighborhood $U$ of $\partial M$, a collar coordinate $\phi \colon \partial M \times [0,\eps) \to U$, and an isomorphism for the graded vector bundle $E$ of fields
\[\phi^*E|_U \iso E_\partial \boxtimes \Lambda^\bullet T^*([0,\eps)),\]
where $E_\partial$ is a graded vector bundle on the boundary such that
the solutions $\Sol_\cT$ to the equations of motion for $\cT$, when viewed as a sheaf on $[0,\eps)$ (i.e., in terms of the collar parameter), is locally constant and takes values in solutions $\Sol_{\cT_\partial}$ for a ``boundary theory''~$\cT_\partial$. 
(We refer to \cite{Rab20} for the full definition of the boundary theory.)
\end{definition}

We will apply Rabinovich's results in the following setting.  
Let $p \colon \mr{Bl}_D(M) \to M$ denote the {\em blow-up} of $M$ along $D$ associated to the tubular neighborhood $U$ and collar coordinate~$\rho$. 
Let's describe it in explicit terms.
Let $D$ have codimension $k$ as a submanifold.
We have fixed a disk bundle $B_3(D)$ inside the normal bundle $N \to D$, equipped with a metric.
Let $S_1(D)$ denote the unit sphere bundle inside $N \to D$.
Observe that there is a natural diffeomorphism
\[
\phi: B_3(D) -D \xto{\cong} S_1(D) \times (0,3)
\]
by using the fiberwise metric;
this map produces a diffeomorphism $U - D \cong S_1(D) \times (0,3)$.
Then
\[
\mr{Bl}_D(M) = (M - D) \cup_{U-D} \left( S_1(D) \times [0,3)\right),
\]
which simply attaches a copy of $D$ to the ``end of the cylinder''~$S_1(D) \times (0,3)$.
The map $p \colon \mr{Bl}_D(M) \to M$ is the identity away from the boundary $\partial \mr{Bl}_D(M) \cong S_1(D) \times \{0\}$, and it collapses the sphere bundle down to~$D$.
Let $\mathring{\mr{Bl}}_D(M)$ denote $\mr{Bl}_D(M) - \partial \mr{Bl}_D(M)$,
and let $\mathring p \colon \mathring{\mr{Bl}}_D(M) \to (M - D)$ denote the restriction of $p$ to the complement of the boundary.

\begin{definition}
For a smooth manifold $M$ with smooth submanifold $D$, let $\cT$ be a classical BV theory on $M - D$.  We say $\cT$ is \emph{topological normal to $D$} if the pullback of $\cT$ along the blow-up map $\mathring p$ extends to a theory on $\mr{Bl}_D(M)$ that is topological normal to its boundary.
\end{definition}

Note a key property of such a theory.
For any choice of radius $0 < r < R < 3$, let $A_{(r,R)}$ denote the annular bundle over~$D$ (i.e., the points in the normal bundle of distance between the radii).
For any choices of radii $0 < r < r' < R' < R < 3$,
the restriction of solutions from the bigger annular bundle $A_{(r,R)}$ to the smaller annular bundle $A_{(r',R')}$ is an equivalence:
\[
\Sol_\cT(A_{(r,R)}) \to \Sol_\cT(A_{(r',R')}).
\]
Thus the observables are likewise equivalent along the extension map
\[
\obscl_\cT(A_{(r',R')}) \to\obscl_\cT(A_{(r,R)})
\]
determined by those radii.  
This claim follows immediately from the topological-normal-to-the-boundary condition,
because the pullback map $\Omega^\bullet((r,R)) \to \Omega^\bullet((r',R'))$ along the inclusion of intervals is a quasi-isomorphism.

Moreover, a choice of local boundary condition $\cL$ on $\mr{Bl}_D(M)$ determines a local boundary condition $\cL_t$ for every~$t$,
as solutions are locally constant with respect to the collar coordinate.

As an immediate corollary of Rabinovich's work, we thus obtain the following result.

\begin{theorem}
If the classical BV theory $\cT$ is topological normal to $D \subset M$,
then a local boundary condition $\cL_t$ determines a factorization algebra $\obscl_t$ on $M$ that is a defect for~$\obscl_\cT$,
i.e.,
\[
\obscl_t|_{M-D} \cong \obscl_\cT|_{M-D}. 
\]
\end{theorem}

The point is that the observables for $\cT$ do not care about the ``width'' of a collar,
so that when we push forward along $\pi_t$,
the stretching of the annular neighborhood is irrelevant.

\begin{remark}
It is important to notice that this theorem applies for a \emph{fixed} value of the radius $t$ around the defect.  This is particular to theories that are topological normal to the boundary.  For theories without this condition, in order to obtain a genuine defect one would need to define a family of theories for all $t$ in the interval $(0,1)$, and take an appropriate limit as $t \to 0$, as discussed briefly in Remark~\ref{limit_remark}.
\end{remark}

\begin{remark}
We expect that the hypothesis can be weakened from ``topological normal to $D$'' to ``rescaling-equivariant normal to $D$.'' 
Compare with \cite{Kapustin}, where Kapustin requires the theory with defect to have a symmetry by the group of conformal transformations preserving the support of the defect.
\end{remark}

Another important feature of Rabinovich's work is that he explains how to quantize classical BV theories that are topological normal to the boundary.
That is, he offers a rigorous renormalization method (building upon Costello's approach in \cite{CosBook} and the work of Albert \cite{Albert}) and formulates a version of the quantum master equation.
His central result is that, when a BV quantization exists (i.e., the master equation is satisfied),
it has a factorization algebra of {\em quantum observables}.
Hence, as another corollary of Rabinovich's work, we have the following. 

\begin{theorem}
If the classical BV theory $\cT$ is topological normal to $D \subset M$ with
a local boundary condition $\cL_t$ that admits a BV quantization,
then the quantum bulk-boundary system determines a factorization algebra $\obsq_t$ on $M$ that is a defect for~$\obsq_\cT$,
i.e.,
\[
\obsq_t|_{M-D} \cong \obsq_\cT|_{M-D}. 
\]
\end{theorem}

\section{Examples}
\label{sec: ex}

We organize our examples by codimension of the defect.
In each example we will focus on how to formulate the field theory and boundary condition along the blowup of a submanifold;
we do not analyze the associated factorization algebra.
At the classical level, most statements about the factorization algebra boil down to statements about the behavior of solutions with the boundary condition.
We postpone such analysis of more interesting examples, including quantizations, to future work.

\begin{remark}
Another rich source of examples arises by applying our construction to examples from \cite{GRW},
which constructs the quantizations and factorization algebras for free bulk-boundary systems (i.e., where the bulk theory is free and the boundary condition is linear).
Thus, the reader can produce defects for abelian Chern-Simons theory (including higher dimensional Chern-Simons theories) or the Poisson sigma-model into a Poisson vector space.
The relevant factorization algebras are analyzed, to some extend, in \cite{GRW}.
See \cite{ButsonYoo} for a number of {\em interacting} bulk-boundary systems of interest to mathematicians and physicists.
We also mention that \cite{CostelloYamazakiIII} contains a wealth of examples and insights about defects that would be well-suited to treatment by the approach advocated by this paper.
\end{remark}

\subsection{Codimension 1}

When the defect's support is a dividing hypersurface, the defect is often called a {\em domain wall}.
We will offer a few examples, building from point defects in mechanical systems towards domain walls for BF theories.

\subsubsection{Topological Mechanics}

Following \cite{GRW, Rab20}, the data of topological mechanics can be encoded in a symplectic vector space~$V$;
in physical terms, we are studying maps from a real line (or worldline) $\RR$ into a target $V$.
The factorization algebra of classical observables for this system encodes the Poisson algebra $\cO(V) = \sym(V^* )$ of functions on~$V$.
A choice of Lagrangian vector subspace $L \subset V$ provides a boundary condition for the half-line $[0, \infty)$; 
in physical terms, it means the path must start in~$L$.
In \cite{GRW}, it is proven that the factorization algebra of classical bulk-boundary observables encodes $\cO(V) = \sym(V^* )$ as the bulk observables together with the module 
\[
\cO(L) = \sym(L^*) \cong \sym((V/L)^*)
\]
as the boundary observables.
In addition, this system can be quantized, 
and \cite{GRW} shows the factorization algebra of quantum bulk-boundary observables encodes the Weyl algebra ${\rm Weyl}(V^*)$ as the bulk observables together with the Fock module ${\rm Fock}(L^*)$  for the boundary observables. 
(As a vector space, ${\rm Fock}(L^*) \cong \cO(L)$.)
The Fock space is a {\em right} module for ${\rm Weyl}(V^*)$ as the boundary is on the left end of the half-line.

We can leverage those results to construct point defects.
Take the point defect for topological mechanics to be supported at the origin,
so the complement of the origin has two boundary points.
Pick Lagrangian subspaces $L_-$ and $L_+$ for boundary conditions on $(-\infty,0]$ and on $[0,\infty)$, respectively.
Our main construction gives a factorization algebra of quantum observables with a defect where on $\RR - \{0\}$, it encodes  ${\rm Weyl}(V^*)$ but for an interval containing the origin,
the observables encode ${\rm Fock}(L_-^*) \otimes {\rm Fock}(L_+^*)$.
Note that this tensor product of Fock modules is a bimodule for the Weyl algebra.

\subsubsection{Domain walls in BF theories} 

Let $\gg$ denote a Lie algebra for a Lie group~$G$. 
Consider BF theory on an oriented $n$-manifold $M$ in the BV formalism: 
the graded vector space of fields is 
\[
\Omega^\sharp(M, \gg)[1] \oplus \Omega^\sharp(M, \gg^*)[2-n]
\]
where an element of the first summand is called the $A$-field
and where an element of the second summand is called the $B$-field.
(We mean here the graded vector space of differential forms, without its differential,
and use $\sharp$ to indicate this.)
The action functional is
\[
S_{\mr{BF}}(A,B) = \int_{M} (B \wedge F_A)
\]
where $F_A = \d A + \frac 1 2 [A,A]$ and the notation $( - \wedge - )$ indicates that we use the evaluation pairing between $\gg$ and its dual $\gg^*$ but wedge the form component.
The equations of motion are $F_A = 0$ and $\nabla_A B = 0$.
In other words, this theory picks out a flat connection and a horizontal section of the coadjoint bundle.

Take $M$ to be a connected manifold and $D$ a dividing hypersurface,
so that
\[
M - D = M_0 \sqcup M_1
\]
is a disjoint union of two manifolds. 
Let $\ol{M}_0 = M - M_1$, and let $\ol{M}_1 = M - M_0$, so that $\ol{M}_i$ is a manifold with boundary isomorphic to $D$.
Then
\[
M \cong \overline{M}_0 \cup_D \overline{M}_1
\]
by construction.

As the equations of motion do not care about the geometry of the manifold, just its underlying topology,
we see that $\Sol_{\mr{BF}}^{\wedge}(D)$, the formal space of solutions near the boundary of $\overline{M}_i$, is modeled by
\[
\Omega^\bullet(D, \gg)[1] \oplus \Omega^\bullet(D, \gg^*)[2-n],
\]
since the de Rham complex in the collar direction is quasi-isomorphic to $\RR$ by the Poincar\'e lemma.

There are many possible choices of local boundary conditions here.
Two obvious options are the summands:
one can take
\[
\Omega^\bullet(D, \gg)[1],
\] 
which corresponds to taking the $B$-field to be zero, or
one can take
\[
\Omega^\bullet(D, \gg^*)[2-n],
\] 
which corresponds to taking the $A$-field to be zero.
Other options arise from taking shifted conormal spaces to a ``subspace'' of~$\Omega^\bullet(D, \gg)[1]$,
which models the formal neighborhood of the trivial connection among all flat $G$-bundles.
(We will mention other possibilities in a moment, to produce ``order'' operators.)

Fix local boundary conditions $\cL_0$ for BF theory on $\overline{M}_0$ and $\cL_1$ for BF theory on $\overline{M}_1$.
Then our main construction produces a factorization algebra on $M$ that agrees with $\obscl_{\mr{BF}}$ on $M - D$ but, for an open set $U$ intersecting $D$, is given by
\[
\obscl_{\cL_0}(U \cap \overline{M}_0) \otimes \obscl_{\cL_1}(U \cap \overline{M}_1).
\]
Note the similarity with the construction in topological mechanics;
by compactifying along $D$ (at least in a collar neighborhood), the two constructions can be identified.

\begin{remark}
\label{rmk: order ops}
We want to remark on another useful class of domain walls, which provide examples of {\em order operators}.
The essential idea is simple: put a field theory $\cT'$ on the hypersurface $D$ that couples to the theory $\cT$ on~$M$.
Away from $D$, the fields of $\cT$ should satisfy the usual equations of motion, 
but on an open intersecting $D$, there is an interesting system of equations involving the fields of both $\cT$ and~$\cT'$.
The observables should thus provide a factorization algebra with a defect along~$D$.
In terms of this paper's point of view, note that if one writes $M$ as the union $(M-U) \cup_{\partial U} U$ for some tubular neighborhood $U$ of $D$,
then one obtains a local boundary condition along $\partial U$ by taking the boundary values of the coupled field theory on~$U$.
(We will not show such boundary conditions are always Lagrangian, but in practice it is typically the case.)  We will discuss examples of this nature in higher codimension in Sections~\ref{CS_fermion_example} and \ref{Wilson_example} below.
An extensive discussion of the order operator case can be found in \cite{PaqWil},
which provides  motivation and many useful examples, as well as references to the pertinent literature.
\end{remark}

\subsection{Codimension 2}

In the setting of BF gauge theories, we exhibit a class of codimension 2 defects that depend upon monodromy of flat connections.
We will then examine point defects for 2-dimensional free scalar theory;
our approach works for point defects of scalar theory in any dimension.
Finally, we  explain how to produce a Wilson line defect for Chern-Simons theory.

\subsubsection{A point defect in 2-dimensional BF theory: monodromy}

\def\mon{{\rm Mon}}

Take $M = \RR^2$ and the origin $D = 0$ as a submanifold.
Let $\gg$ be the Lie algebra of a Lie group~$G$.
If we consider BF theory for $\gg$ on $\mathring{M} = \RR^2-\{0\}$,
then a solution to the equations of motion provides a flat connection $\nabla = \d + A$.
Such a flat connection has monodromy $\mon(\nabla)$ for a loop $\gamma$ that winds once around the origin,
and this monodromy is an element of $G$, only up to conjugation.
(And every element of $G$ can appear as monodromy. )
In other words, we have $\mon(\nabla) \in G/G$, the adjoint quotient space,
and, in fact, two flat connections with the same monodromy are gauge-equivalent (for a textbook account, see \cite[\S 13.2]{Taubes}).

In a BF theory, a solution also involves a choice of $\nabla$-closed $\gg^*$-valued function $B$,
i.e., a $\gg^*$-valued function that is horizontal for~$\nabla$.
Such a function is completely determined by its value at one point $p$ in $\mathring{M}$,
i.e., an element of $\gg^*$.
Thus, the quotient space $(G \times \gg^*)/G$, with respect to the adjoint-coadjoint action, parametrizes solutions to the BF theory on $\mathring{M}$,
and hence perturbative BF theories.
Given a pair $(m, b) \in G \times \gg^*$, let $\cT_{(m,b)}$ denote the perturbative BF theory around a pair $(\nabla, B)$ lying in the corresponding $G$-orbit, i.e.
\[[\mon(\nabla), B(p)] = [m,b].\]
Let $\obscl_{(m,b)}$ denote the factorization algebra on $\mathring M$ of classical observables of this theory.

We note that such a theory is topological.  In particular, if we choose any closed disk $D_t$ of radius $t < |p|$, the formal space of jets of solutions $\Sol_{\mr{BF}}^{\wedge}(\partial D_t)$ at the boundary circle $\partial D_t = \partial (\RR^2 - \mathring D_t)$ is canonically independent of $t$.

Suppose for simplicity that $[m,b] = [e,0]$.  Similarly to the previous example, this space of formal jets of solutions is modeled by the shifted de Rham complex
\[
\Omega^\bullet(\partial D_t, \gg \oplus \gg^*)[1].
\]
If we choose a non-trivial value of $[m,b]$ then we must instead use a twisted de Rham complex with differential modified by the chosen basepoint.  We can choose natural boundary conditions by taking forms valued in any Lagrangian linear subspace of $\gg \oplus \gg^* = T^*\gg$.  There are several natural classes of example.
\begin{enumerate}
 \item If $\gg$ admits an inner product $\kappa \colon \sym^2(\gg) \to \RR$, there is an induced linear isomorphism $f_\kappa \colon \gg \to \gg^*$, and there are boundary conditions $\cL_s$ associated to each real number $s$, given by the image of the map $(1,sf_\kappa) \colon \gg \to \gg \oplus \gg^*$.  Such boundary conditions impose the condition $B = sf_\kappa(A)$ at the boundary.
 
 \item If we choose a Lie subalgebra $\mf l \sub \gg$, there is an associated Lagrangian subspace $\mf l \oplus (\mf l^*)^\perp \sub \gg \oplus \gg^*$, inducing associated boundary conditions $\cL_{\mf l}$.  Such boundary conditions restrict the monodromy of $\nabla$ to lie in the subalgebra $\mf l$.
\end{enumerate}

\begin{remark}
There was nothing particular to $M = \RR^2$ in the calculation we just outlined.  If we take $M = \RR^n$ and take $D$ to be a choice of embedding $\RR^{n-2} \to M$, then we can perform an identical calculation.  For example, this provides a description of a class of line defects in 3-dimensional BF theory, and of a class of surface defects in 4-dimensional BF theory.
Small and straightforward modifications allow one to construct such ``monodromy defects'' for a codimension 2 submanifold inside any oriented manifold,
although characterizing the allowable monodromies can be elaborate.
\end{remark}

\begin{remark}
As we discussed in Section \ref{rmk on history}, each choice of boundary condition will lead to a module for the algebra of classical observables on an annulus.  One can describe this algebra fairly concretely starting from the dg Lie algebra $\Sol_{\mr{BF}}^{\wedge}(\partial D_t)[-1]$, by forming the Chevalley--Eilenberg cochains.
\end{remark}

\subsubsection{Point defects in scalar field theory}

We will now move on to discussing another example of point defects in $M = \RR^2$, 
but now in a non-topological theory (so the results of Section \ref{sec: theorem} will not be applicable).  
We will consider the example of a free classical scalar field theory on $\RR^2$ equipped with its flat metric, with a defect at the origin, so $D = \{0\} \sub \RR^2$.  
The blow-up is easy to describe here: it is given by excising a disk $D_R(0)$, so 
\[ \RR^2 - D_R(0) = \{z \in \RR^2 \colon |z| \ge R\}.\]
We will trace the dependence on radius $R$ below.

The free scalar field theory has graded space of fields $$C^\infty_\CC(\RR^2) \oplus C^\infty_\CC(\RR^2)[-1],$$ 
where $C^\infty_\CC(\RR^2)$ denotes the space of smooth {\em complex}-valued functions.
The action functional is
\[S(\phi) = \int \phi \Delta \phi\]
and depends only on the field in cohomological degree zero.  Here $\Delta$ is the usual Laplacian operator.  The formal space of solutions near $\phi = 0$ is modelled by the complex
\[C^\infty_\CC(\RR^2) \overset {\Delta}{\to} C^\infty_\CC(\RR^2)\]
concentrated in degrees zero and one.  The space of jets of solutions along the boundary $S^1_R$ of the blowup can also easily be described.  Let us use polar coordinates $(r,\theta)$ on $\RR^2$, and let $t = r-R$ be a collar coordinate near our boundary.  Then jets of fields near the origin are modelled by 
\[C^\infty_\CC(\RR^2)^\wedge = C^\infty_\CC(S^1)[[t]],\]
and the differential becomes
\begin{align*}
\Delta^\wedge &= \partial^2_t + \frac 1{t+R}\partial_t + \frac 1{(t+R)^2}\partial_\theta^2 \\
&= \partial^2_t + \frac 1R \sum_{m \ge 0}\left(\frac tR\right)^m \partial_t + \frac 1{R^2} \left(\sum_{m \ge 0}\left(\frac tR\right)^m\right)^2 \partial_\theta^2.
\end{align*}
We will use the more attractive $\mathcal{D}$ to denote this operator $\Delta^\wedge$ obtained by restricting to jets along the boundary $S^1_R$,
and we write it succinctly as
\[\mathcal{D}=\partial_t^2+g_R(t)\partial_t+g_R(t)^2\partial_{\theta}^2,\]
where \[g_R(t)= \frac 1 R \sum_{m \geq 0} \left ( \frac t R \right)^m \] denotes the geometric series. 
It is routine to verify the following properties of the operator $\mathcal{D}$:
\begin{enumerate}
\item [(1)] $\mathcal{D}$ is surjective and thus $\coker(\mathcal{D})=0$.
\item [(2)] $\ker(\mathcal{D}) \cong C^{\infty}_\CC(S^1)$.
\end{enumerate}
(This second fact boils down to showing that any solution is determined by its $t^0$-term.)
These facts imply that there is a quasi-isomorphism of the form
\[C^{\infty}_\CC(S^1)\hookrightarrow \Sol^{\wedge}_{\partial},\]
where the injective map is defined by the inclusion of the kernel of the operator $\mathcal{D}$.  This inclusion is a section of a quasi-isomorphism $\Sol^{\wedge}_{\partial} \to C^{\infty}_\CC(S^1)$ given by projection onto the $t^0$ component in the degree zero part of~$\Sol^{\wedge}$.
We thus have a concrete description of boundary conditions for the free scalar field on $\RR^2$. 

We observe that there is a natural presymplectic structure on $\ker \mathcal{D} \cong \mathcal C^{\infty}_\CC(S^1)$ given by
\[ \omega_D(e^{ik\theta}, e^{il\theta})=
\begin{cases}
        1 & \text{if } k>0, l=-k, \\
        -1& \text{if } k<0, l=-k, \\
        0& \text{if } k\neq l.
    \end{cases}
\]
Note that this is well-defined as a pairing on smooth functions on the space $S^1$ because the Fourier coefficients of any smooth function decay more rapidly than any polynomial.

This presymplectic structure is degenerate due to the existence of the constant Fourier coefficient.  As a result, our example only satisfies a weakened version of Hypothesis~\ref{symplectic_hypothesis}.  
We could resolve this issue by a mild alteration, where we replace the complex controlling jets of solutions by its quotient by the constant factor~$\CC$.

The following are examples of Lagrangians  $L \Longrightarrow C^{\infty}_\CC(S^1)/\CC$ with respect to this symplectic form.
\begin{enumerate}
\item [(a)] Take the positive modes
\[L_0 = \{f \in C^\infty_\CC(S^1) \colon \hat f_n = 0 \text{ if } n < 0\}/\CC.\]
Here $\hat f_n$ denote the $n^{\text{th}}$ Fourier coefficient of the function $f$.  This boundary condition corresponds to the inclusion of the space of functions that extend across the interior disk, so it is the ``trivial defect" that recovers the scalar field theory on~$\RR^2$.

\item [(b)] More generally, we could consider the following Lagrangian:
\[L_0 = \{f \in C^\infty_\CC(S^1) \colon \hat f_n = 0 \text{ if } n \in S\}/\CC.\]
 where $S \subset \ZZ$ is a subset such that $0 \notin S$ and $k \in S \iff -k \notin S$, for all $k \ne 0$.  We refer to such Lagrangians as ``spectral" boundary conditions as they are determined by spectral properties of the function along the boundary and not pointwise behavior along the boundary.
\end{enumerate}
A variation of (a) is the following: if $\eta$ denotes the choice of a conformal class of a metric on $D_R(0)$, then 
\[L_{\eta}= \left \{ \eta \mbox{--harmonic functions on the disk} \right \} \]
is a Lagrangian.
(This construction offers a useful view on the appearance of the ``restricted" or ``semi-infinite" Grassmannian in conformal field theory.)One can formulate more exotic spectral boundary conditions too.

\begin{remark}
We point out that our approach here works in higher dimensions. 
For free scalar theory on $\RR^n$ with a point defect at the origin,
one simply replaces $S^1$ by $S^{n-1}$,
and the Fourier modes by spherical harmonics.
The remaining analysis is parallel.
\end{remark}
 
\subsubsection{Chern--Simons theory coupled to a charged fermion: a Wilson line} 
\label{CS_fermion_example}

Let us now take $M = \RR^3$ and $D = \RR \times (0,0)$, the $x$-axis.  We will describe an ``order'' type defect obtained by coupling a background gauge theory on $M$ to a charged particle along the line $D$.  Thus, fix a Lie algebra $\gg$ with a non-degenerate pairing, and let $V$ be a finite-dimensional representation of $\gg$ equipped with an invariant inner product.  We define our BV  theory on $\RR^3$ to be \[\cT = \Omega^\bullet(\RR^3) \otimes \gg[1],\]
with the BV bracket defined using the pairing on $\gg$.  This is a perturbative description of \emph{Chern--Simons theory} near the trivial flat connection.

If we choose a tubular neighborhood $U$ of the embedded line $\RR$, with boundary $\partial U$, the formal space $\Sol^\wedge_{\mr{CS}}(\partial U)$ of jets of solutions near the boundary is determined by the shifted de Rham complex
\[\Omega^\bullet(\partial U) \otimes \gg[1],\]
with cohomology concentrated in degrees $-1$ and $0$.

We can define a boundary condition in a trivial way.  
\begin{definition}
Let $i_\partial \colon \partial U \to U$ denote the inclusion of the boundary of $U$.  There is a canonical Lagrangian given by the restriction map
\[i_\partial^*(\Omega^\bullet(U) \otimes \gg[1]) \to \Omega^\bullet(\partial U) \otimes \gg[1].\]
It is the \emph{trivial} defect along $D$.
\end{definition}

We can enhance this trivial defect by coupling to an additional field along the line $D$, valued in the representation $V$.  Define a topological \emph{free fermion} theory on $\RR$ by setting
\[\cT_{V} = \Omega^\bullet(\RR) \otimes \Pi V\]
with BV pairing defined using the inner product on $V$ and wedge-and-integration of the forms.
Here $\Pi$ indicates that we place the representation $V$ in odd degree for an auxiliary $\ZZ/2\ZZ$-grading.  
The action functional is
\[
S_V(\psi) = \int_\RR (\psi, \d \psi)
\]
for the associated BV field theory, with $\psi \in \cT_V$.

We will now define a defect by coupling the topological free fermion $\cT_V$ along $D$ to the Chern-Simons theory in the bulk.  
Observe that the $\gg$-module structure on $V$ makes $\cT_{V}|_U[-1]$ into a module for the sheaf of dg Lie algebras $i^*_D\cT[-1]$, where $i_D$ is the inclusion of the defect line in~$U$.
In terms of action functionals, this means that we can view the gauge field $A$ as modifying the fermion action by minimal coupling:
\[
S_{V,{\rm min}}(\psi; A) = \int_\RR (\psi, (\d + A)\psi).
\]
The equation of motion for this theory picks out sections $\psi$ that are horizontal (or flat) for the connection $\d + A$;
if we worked with a circular defect rather than a line, we could ask about the holonomy of the connection on this vector bundle.
In this way, the fermionic theory allows one to encodes the Wilson operator (i.e., trace of holonomy).

We can formulate a BV theory that involves both the gauge field and the fermion in terms of the super dg Lie algebra 
\[
\mc E_{\gg,V}(U) = \left(\Omega^\bullet(U) \otimes \gg\right) \ltimes (i_D)_*\left(\Omega^\bullet(\RR) \otimes \Pi V[-1]\right)
\]
with bracket generated by the Lie bracket on $\gg$ and the action of $\gg$ on~$V$.
The associated action functional is the sum of the Chern-Simons action and minimally coupled action for the fermion.

\begin{definition}
The {\em charged line defect} associated to a free fermion valued in the representation $V$ of $\gg$ is the boundary condition
\[i_\partial^* \mc E_{\gg,V}(U) \to \Omega^\bullet(\partial U) \otimes \gg[1].\]
\end{definition}

If we set $V = 0$, we recover the trivial defect defined above.

This charged line defect is another example of an {\em order} operator, as discussed in Remark~\ref{rmk: order ops}.

\begin{remark}
Note that once we introduce a non-trivial representation $V$, 
the boundary condition is no longer associated to the inclusion of a subcomplex.
In the derived setting, as we have here, a map can admit a Lagrangian structure even if it is not a degreewise inclusion.
\end{remark}

\begin{remark}
Let us conclude this section by mentioning an additional interesting collection of codimension two defects that one could hope to describe in the present formalism.  
Costello and Yamazaki \cite{CostelloYamazakiIII} consider a large class of surface defects of both order and disorder type in \emph{four-dimensional Chern--Simons theory}: a gauge theory analogous to Chern--Simons theory, but defined on $\RR^2 \times C$ where $C$ is a Riemann surface, in which the solutions to the equations of motion are topological in the two real directions and holomorphic in the complex direction.  
They consider defects placed along planes of the form $\RR^2 \times \{z\}$ for points $z \in C$, from which they are able to engineer a large number of interesting integrable systems on the Riemann surface.  We should note that these examples will not be topological normal to the boundary, but only holomorphic in the normal directions.
\end{remark}

\subsection{Codimension 3}

We now turn to the most well-known examples from gauge theory:
we describe the magnetic monopole and the Wilson line in 4-dimensional Yang-Mills theory.  The results of Section \ref{sec: theorem} will not apply in this section, because Yang--Mills theory is not topological (it may be possible to recover similar results using the weaker condition of conformal invariance, but do not make any claims in this direction at present).

\subsubsection{The magnetic monopole}

On four-dimensional manifolds, Yang--Mills theory admits a first-order formulation that is convenient for producing boundary conditions,
so we will stick to dimension~4 in this paper.
We will also only consider Yang--Mills theory for the abelian group~$\mr U(1)$.
(In a companion paper \cite{FAofMonopole}, we will work out the monopole for abelian Yang--Mills theory in arbitrary dimensions.)

Let $M$ be a Riemannian 4-manifold with Hodge star operator $\star$.
(In fact, to describe a Yang--Mills theory, we only need the data of a Hodge star operator---up to rescaling---not the full metric.)
Fix a principal $\mr U(1)$-bundle $L \to M$ with a non-degenerate inner product, and fix a connection $\nabla$ on~$L$.
(The choice allows us to describe the space of connections as sections of a vector bundle.)
The first-order formulation of $\mr U(1)$-Yang--Mills theory has graded space of fields
\[
\begin{array}{cccc}
-1 & 0 & 1 & 2 \\
\hline
\Omega^0(M, L) & \Omega^1(M, L) & \Omega^2_+(M, L) & \\
 & \Omega^2_+(M, L) & \Omega^3(M, L) & \Omega^4(M, L) 
 \end{array}
\]
where $\Omega^2_+(M, L)$ denotes the self-dual 2-forms,( i.e., $\alpha$ such that $\star \alpha = \alpha$, using the inner product to identify $L$- and $L^*$-valued 2-forms)
and where the top row indicates the cohomological degree.
We call an element of the first row the $A$-field and an element of the second row the $B$-field.
The action functional is
\[
S_{\mr{YM}}(A,B) = \int_M B \wedge \nabla_+ A - c (\star B) \wedge A
\]
where $c$ denotes a coupling constant and where $\nabla_+$ denotes the covariant derivative followed by projection onto self-dual 2-forms.
The equations of motion are 
\[
\nabla_+ A = c B \qquad \text{and} \qquad \nabla B = 0,
\]
which together imply the usual equations of motion.  For instance if $c \ne 0$, we deduce that $\nabla \ast \nabla A = 0$. (When $c = 0$, we recover self-dual Yang--Mills theory.)

A solution to the equation of motion for $\mr U(1)$-Yang--Mills theory is a connection $\wt \nabla = \nabla + A$ on the bundle~$L$.
The tangent complex at $\wt \nabla$ to the derived space of solutions is modelled by the cochain complex
\begin{equation} \label{YM_complex}
\mc E_M =\quad 
\vcenter{\hbox{\xymatrix{
\Omega^0(M,L) \ar[r]^{\wt \nabla} &\Omega^1(M,L) \ar[r]^{\wt \nabla_+}  &  \Omega^2_+(M,L) & \\
& \Omega^2_+(M,L) \ar[r]^{\wt \nabla} \ar[ur]^{-c \cdot \star} & \Omega^3(M,L) \ar[r]^{\wt \nabla} & \Omega^4(M,L)
}}}\end{equation}
concentrated in degrees $-1, 0, 1, 2$. Note that there is a natural subcomplex given by the top row, i.e., the subcomplex consisting only of $A$-fields.
If $M$ is a manifold with boundary, then jets of this subcomplex along $\partial M$ yield a local boundary condition.
Let $\cL_{B = 0}$ denote this boundary condition. 

\begin{remark} \label{YM_boundary_remark}
We can describe this in a little more detail, though we will not include a detailed calculation in this paper.  We can model the jets of solutions along $\partial M$, i.e. the formal space $\Sol^\wedge_{\mr{YM}}(\partial M)$, by the cochain complex
\begin{equation} \label{YM_boundary_complex}
\mc E_{\partial M} = \quad 
\vcenter{\hbox{\xymatrix{
\Omega^0(\partial M,L|_{\partial M}) \ar[r]^{\wt \nabla|_{\partial M}} &\Omega^1(\partial M,L|_{\partial M}) & \\
&z\Omega^2(\partial M,L|_{\partial M}) \ar[r]^{\wt \nabla|_{\partial M}} &z\Omega^3(\partial M,L|_{\partial M})
}}}\end{equation}
concentrated in degrees $-1, 0$ and 1, where $z$ is a chosen formal coordinate normal to $\partial M \sub M$.  So $\cL_{B=0}$ is equivalent to the complex
\[\xymatrix{
\Omega^0(\partial M,L|_{\partial M}) \ar[r]^{\wt \nabla|_{\partial M}} &\Omega^1(\partial M,L|_{\partial M})}\]
in degrees $-1$ and 0.
\end{remark}

We can now finally formulate magnetic monopoles, as follows.
Let $M = \RR^4$ and fix a line $D \subset M$, so $D \cong \RR$.
We require that $D$ ``extend to infinity,'' i.e., it is not contained in any compact subset of~$M$;
for simplicity, we also require that $D$ is unknotted.
For example, when the metric is Euclidean, use an axis of the usual Cartesian coordinate system.
The manifold $M - D$ is then diffeomorphic to $S^2 \times \RR^2$.
In this situation, $\mr U(1)$-bundles up to isomorphism are computed by $\mr H^1(M-D, \mr U(1)) \cong \mr H^2(M-D, \ZZ)$, which is isomorphic to $\ZZ$.
For a line bundle $L$, its first Chern class $c_1(L)$ is the identifying cohomology class.
Any line bundle $L$ does admit solutions to the equations of motion for $\mr U(1)$-Yang--Mills theory,
and a physicist would call $c_1(L)$ the {\em magnetic charge} of the solution.
When $c_1(L) \neq 0$, such a solution is called a magnetic monopole.
(We can view the case of $c_1(L)=0$ as a charge-free monopole.)

Thus, let $\nabla$ be a magnetic monopole with magnetic charge~$m$.
Then if we impose the boundary condition $\cL_{B = 0}$ along the line $D$, 
we can use our construction for any small value of $t$ to produce an effective factorization algebra for the monopole.

\begin{remark}
 There is a qualitatively different story describing monopoles in non-abelian Yang--Mills--Higgs theory (such as 't Hooft--Polyakov monopoles), where there is no locus on which the fields become singular.  We will discuss such examples in~\cite{FAofMonopole}.
\end{remark}

\subsubsection{Wilson line defects in Yang--Mills theory} 
\label{Wilson_example}

As a final example, let us discuss Wilson lines in Yang--Mills theory. 
This example will be defined very similarly to the order type operators described above, 
particularly in Section~\ref{CS_fermion_example}.  

To begin, recall that the irreducible representations of $\mr U(1)$ are all one-dimensional, labelled by their weight $n \in \ZZ$.  We will denote this representation by $V_n$; 
we will view $n$ as the {\em electric charge} of a charged particle.  

Let $M = \RR^4$, and let $D \sub M$ be an embedding of a line with tubular neighborhood $U$.  Write $i_D \colon D \to U$ for the inclusion.  We can describe a ``coupled'' theory along the line using the complex
\[\mc E_{U,n} = \mc E_U \ltimes (i_D)_*(\Omega^\bullet(D) \otimes V_n),\]
where $\mc E_U$ is the complex from Equation 
\eqref{YM_complex}, and where $\mf u(1) \ltimes V_n$ is the Lie algebra with underlying vector space $\RR \oplus V_n$ whose only non-trivial bracket is generated by 
\[[(1, 0), (0, v)] = (0, n v).\]
This is an example of the representations $\gg \ltimes V$ described in Section~\ref{CS_fermion_example}, but with the representation $V$ now placed in even rather than odd degree.

We can define the Wilson line defect associated to the representation $V_n$.
\begin{definition}
Choose a tubular neighborhood $U$ of $D$ in $M$, and let $\iota_\partial$ be the embedding of the boundary of $U$ in $M$.  The \emph{Wilson line defect} with electric charge $n$ is the defect
\[\iota_\partial^* \mc E_{U,n} \to \mc E_{\partial U},\]
where $\mc E_{\partial U}$ is the complex modelling germs of solutions to the equations of motion along $\partial U$ (e.g. as defined in Equation~\eqref{YM_boundary_complex}).
\end{definition}

\begin{remark}
If we \emph{also} chose a magnetic monopole with magnetic charge $m$, singular along $D$, we could similarly define a Wilson line defect with electric charge $n$ in this background: one can define a defect by taking the magnetic charge $m$ monopole defect $\mc L_{B=0}$ and forming the tensor product with the electric charge $n$ Lie algebra $\mf u(1) \ltimes V_n$ defined above.  The result is called a \emph{dyonic defect} with charge~$(m,n)$. 
\end{remark}

\printbibliography
\end{document}